\documentclass[12pt,preprint]{aastex}

\shorttitle{The crab flare }
\shortauthors{Teraki et al.}

\begin{document}

\title{Jitter radiation model of the Crab gamma  ray flares }

\author{Yuto Teraki\altaffilmark{1} and Fumio Takahara\altaffilmark{1}}
\affil{Department of Earth and Space Science, Graduate School of Science, 
Osaka University, 1-1 Machikaneyama-cho, Toyonaka, Osaka 560-0043, Japan}
\email{teraki@vega.ess.sci.osaka-u.ac.jp}

\begin{abstract}
The gamma ray flares of the Crab nebula detected by Fermi and AGILE satellites challenge
our understanding of physics of pulsars and their nebulae.
The central problem is that the peak energy of the flares exceeds the maximum energy $E_{\mathrm{c}}$ 
determined by synchrotron radiation loss.
However, when there exist turbulent magnetic fields with scales $\lambda_{\mathrm{B}}$ smaller than $2\pi mc^2/eB$, 
 jitter radiation can emit photons with energy higher than $E_{\mathrm{c}}$.
The scale required for the Crab flares is about two orders of magnitude less than the wavelength of the striped wind.
We discuss the model in which the flares are triggered by plunging of the high density blobs into the termination shock.
The observed hard spectral shape may be explained by jitter mechanism.
We make three observational predictions: firstly the polarization degree will become lower in flares, secondly, no counterpart will be seen in TeV-PeV range,
and thirdly the flare spectrum will not be harder than $\nu F_\nu \propto \nu^1$.
\end{abstract}


\section{Introduction}
The Crab nebula has been known as a luminous celestial object and has been regarded as a stationary emitter except for a secular change due to the expansion.
Recently, strong flares were detected five times in the range $>100$ MeV by AGILE (Tavani et al. 2011) and Fermi (Abdo et al. 2011, Buehler et al. 2012, Ojha et al. 2012) satellites.
The flares occur about once in a half year, the flux doubling timescale is around 8 hours, and duration time is a few weeks.
The peak energy is as high as $375\mathrm{MeV}$ which is a challenge for the standard scenario of pulsar wind nebulae (Buehler et al. 2012).
When electrons/positrons are accelerated on gyro timescale, synchrotron radiation limits the attainable energy 
(see e.g.| Kirk \& Reville 2010), and
the maximum energy of synchrotron radiation is $\sim 100\mathrm{MeV}$. 
Since there seem to be no counterparts in other energy ranges, they should involve only the highest energy particles.
In fact, the flare of April 2011 shows very hard spectrum with
the photon index $\gamma_{\mathrm{F}} = 1.27 \pm 0.12$ (Buehler et al. 2012).
The peak flux amounts to $(186 \pm 6) \times 10^{-7} \mathrm{cm}^{-2}\mathrm{s}^{-1}$, 30 times larger than the quiescent one.
The isotropic luminosity amounts to $4 \times 10^{36} \mathrm{erg}\: \mathrm{s^{-1}}$ corresponding to $1\%$
of the spin down luminosity of the Crab pulsar.
The size of emission region of the flares should be as small as $ct_{\mathrm{fluc}} \sim 10^{15}\mathrm{cm}$ or $ct_{\mathrm{dur}} \sim 3\times 10^{16} \mathrm{cm}$,
where $t_{\mathrm{fluc}}$ is the fluctuation timescale estimated from flux changes and $t_{\mathrm{dur}}$ is the duration timescale of the flares. 
Either of them is very small compared to the circumference of the termination shock $2\pi r_{\mathrm{ts}}\sim 2 \times 10^{18} \mathrm{cm},$
 where $r_{\mathrm{ts}}$ is the radius of the termination shock from the Crab pulsar.
It is notable that such a large amount of energy is concentrated in a small region.

Although several models have been proposed to overcome the crucial problem of $E_{\mathrm{c}}$, the consensus has not been achieved.
The obvious possibility is relativistic beaming effect.
In the standard scenario of pulsar wind nebulae (e.g. Kennel \& Coroniti 1984), the bulk speed of nebula region is nonrelativistic,
but a possibility of the emission regions having relativistic speed was discussed from various aspects.
(Komissarov \& Lyutikov 2011, Bednarek \& Idec 2011, Yuan et al. 2011, Kohri et al. 2012, Clausen-Brown \& Lyutikov 2012).
Another possibility is a separation between the acceleration region and emission region (Uzdensky et al. 2011, Cerutti et al. 2012).
They considered the acceleration by the electric field on a reconnection sheet.
The magnetic field on the reconnection sheet is much weaker than outside the sheet, and electrons can be accelerated
by the electric field suffering from much weaker radiation loss and achieve a larger Lorentz factor.
In somewhat different view point, Bykov et al. (2012) considered effects of inhomogeneities of the magnetic field strength. 
The highest Lorentz factor of electrons is limited by the mean strength of magnetic field,
and the highest energy emission comes from small regions where the magnetic field is strongest.
The spatial scale of the acceleration region is the same order of the Larmor radius of the highest energy electrons
$r_{\mathrm{L}} \sim  2 \times 10^{17} \left(\frac{\gamma}{10^{10}}\right) \left(\frac{B}{10^{-4}\mathrm{G}}\right)^{-1} \mathrm{cm}$,
while the scale of the emission region is as small as $c t_{\mathrm{fluc}} \sim 10^{15}\mathrm{cm}$ or $c t_{\mathrm{dur}} \sim 3 \times 10^{16} \mathrm{cm}$.
If the magnetic field varies by a factor of $3$ in a small region,
the emission energy can be higher than $E_{\mathrm{c}}$ in this case.

A common feature of these models is that the radiation process is considered to be synchrotron radiation.
In contrast, we consider yet another possibility that the magnetic fields become turbulent on very small scales, and
radiation process changes from synchrotron radiation to jitter radiation.
The photon energy of jitter radiation can be higher than $E_\mathrm{c}$ in this situation (Fleishman 2006).
For the jitter radiation, the typical frequency is determined by the scale $\lambda_{\mathrm{B}}$ of the turbulent magnetic field.
We suppose that this scale is much smaller than $2 \pi mc^2/eB$ and that the electrons move approximately straightly.
The typical frequency is $\gamma^2$ times the inverse of the timescale that the electrons move across $\lambda_{\mathrm{B}}$, and
\begin{equation}
 \omega_{\mathrm{B}} \sim \gamma^2 2\pi c/\lambda_{\mathrm{B}}.
\end{equation}
Therefore, photons with frequencies higher than $\gamma^2 eB/mc$ can be emitted
if the spatial scale of the turbulent magnetic field is smaller than $2\pi mc^2/eB$.

In this paper, we discuss the flare model based on jitter radiation.
In section 2, we explain possibilities to create the flares by jitter radiation and discuss the flare energetics and spectra.
In section 3, we discuss differences from other models.
We summarize this paper in section 4.

\section{Jitter radiation}
\subsection{Small scale turbulence}\label{sec:small}
The jitter radiation is the radiation from a relativistic particle moving in a random magnetic field with the spatial coherence scale
shorter than the typical synchrotron photon formation length (Medvedev 2000).
We assume that the turbulent magnetic field is isotropic in this paper. 
When $\lambda_{\mathrm{B}} < 2 \pi mc^2/eB$, in other words, when the strength parameter
\begin{equation} 
a \equiv \frac{eB\lambda_{\mathrm{B}}}{2 \pi mc^2}
\end{equation}
 is smaller than 1, jitter approximation is valid (Medvedev et al. 2011, Teraki \& Takahara 2011).
Using the condition for jitter approximation of $a<1$, we can write the strength of magnetic field of the emission region as
\begin{equation}
B < 1 \times 10^{-3}(\frac{\lambda_\mathrm{B}}{10^7\mathrm{cm}})^{-1}\mathrm{G}. \label{eq:b}
\end{equation}
We suppose that the acceleration site for the flares is near the shock front.
We tentatively assume that the magnetic field becomes turbulent in a small part of the acceleration region, 
though we consider later that the size of them are same order.
The Lorentz factor of accelerated electrons which emit the highest energy synchrotron photons $\sim 100\mathrm{MeV}$
in a quiescent state is thought to be $\sim 10^{10}$ and the average magnetic field strength of $\sim 10^{-4}\mathrm{G}$
(Kennel \& Coroniti 1984, De Jager \& Harding 1992, Atoyan \& Aharonian 1996, Tanaka \& Takahara 2010).
The required scale of turbulence to meet the condition $a<1$ is $\lambda_{\mathrm{B}} <10^8\mathrm{cm}$ when magnetic field strength is $10^{-4}\mathrm{G}$.
On the other hand, the required scale to emit flare photons with energy $\sim 400\mathrm{MeV}$ by the highest energy electrons through the jitter radiation,
the required scale of turbulent magnetic field is $\sim 3\times 10^7\mathrm{cm}$.

We note that the wavelength of the striped wind of the Crab pulsar
 ($\lambda_{\mathrm{sw}} \equiv c \times 33\mathrm{ms} \sim 10^9 \mathrm{cm}$) is around the required length.
Our picture of the flares is expressed as follows.
When alternating magnetic fields are injected into the acceleration site, fluctuations with scales shorter than $\lambda_{\mathrm{sw}}$
are generated through compression or transformation to some type of waves.
The highest energy electrons feel the small scale magnetic fields, and radiate high energy photons by jitter mechanism.
In the quiescent state this mechanism may not work, because the density in the pulsar wind is very low, and
the small scale turbulent field is suppressed, as we see in the next paragraph.
Here we consider here how the small scale magnetic field can be generated when the flares occur.
In general, the pulsar wind fluctuates temporarily and spatially.
For example, the Crab pulsar is known to emit very energetic radio pulses, called "Giant Radio Pulse" (GRP) about once in thousands 
(e.g. Lundgren et al. 1995).
This suggests that there may be large density fluctuations in the magnetosphere.
Furthermore, from the observations of these GRPs, it has been argued that the dispersion measure fluctuates largely,
and these fluctuations can not be explained
by considering the density fluctuations of the interstellar medium alone.
Therefore, it is suggested that there are large density fluctuations in the Crab nebula (Kuz'min et al. 2008, 2011).
From these observations, it is quite natural to suppose that there are density fluctuations in the wind region.
We advocate the model that plunging of a high density blob into the termination shock triggers a flare.
We note, however, that the flares are not directly the same events as GRP (Mickaliger et al. 2012).

Next we compare the wavelength of striped wind and the typical scales of plasma in the comoving frame,
and consider the conditions for survival of small scale magnetic fields. 
Although the striped wind itself is a non-propagating entropy mode, existence of high density blobs and moderate reconnection may generate
electrostatic and electromagnetic modes on somewhat shorter wavelength than $\lambda_{\mathrm{sw}}$.
We may consider various modes, for example, electron Bernstein mode, which is the electrostatic wave in a thermal plasma (Bernstein 1958),
but we do not specify the type of plasma turbulence.
When the inertial length is longer than the $\lambda_{\mathrm{sw}}$, the electromagnetic modes can survive, while the short scale electrostatic mode may decay.
To estimate the typical scale of the survival of the longitudinal modes, we use the value of inertial length.
First we consider it in the upstream, i.e., wind region.
The Debye length is very small compared to the inertial length, because the plasma is cold when the reconnection is moderate.
The inertial length $c/\omega_{\mathrm{pe}}$ can be estimated given the comoving number density.
The spindown luminosity is expressed by
\begin{equation}
L_{\mathrm{sd}} = 4 \pi r_{\mathrm{ts}}^2 n \Gamma u mc^3(1+\sigma) = 6 \times 10^{38}\mathrm{erg}{\mathrm{s}^{-1}}, \label{eq:ef-cons}
\end{equation}
where $r_{\mathrm{ts}} = 3 \times 10^{17}\mathrm{cm}$, $n$ is the comoving number density, $\Gamma=10^6$ (Kennel \& Coroniti 1984)
or $\Gamma = 7 \times 10^{3}$ (Tanaka \& Takahara 2010)
is the bulk Lorentz factor of the pulsar wind, $u$ is the radial four velocity, $\sigma$ is the ratio of magnetic to kinetic energy flux.
In general, $\sigma$ is thought to be much smaller than 1 at the shock region
( $\sigma \sim 0.003$ is the best fit value in Kennel \& Coroniti 1984).
We adopt this assumption, and neglect $\sigma$ in (\ref{eq:ef-cons}).
When we adopt the value of the bulk Lorentz factor by Tanaka \& Takahara 2010,
we get the comoving density $n \sim 4 \times 10^{-10}\mathrm{cm}^{-3}$ and the value of inertial length
\begin{equation}
\left( \frac{c}{\omega_{\mathrm{pe}}}\right)_\mathrm{u,TT} \sim 3 \times 10^{10}\mathrm{cm}.
\end{equation}
When we adopt $\Gamma = 10^6$ (Kennel \& Coroniti model), the comoving density becomes smaller.
Using the equation (\ref{eq:ef-cons}), we get $n \sim 2 \times 10^{-14}\mathrm{cm}^{-3}$, and we obtain
\begin{equation}
\left(\frac{c}{\omega_{\mathrm{pe}}}\right)_\mathrm{u,KC} \sim 3 \times 10^{12}\mathrm{cm}.
\end{equation}
On the other hand, the comoving wavelength of striped wind is
\begin{equation}
(\Gamma \lambda_{\mathrm{sw}})_{TT} \sim 1\times 10^{12}\mathrm{cm},\/ 
\end{equation}
\begin{equation}
(\Gamma \lambda_{\mathrm{sw}})_{KC} \sim 2 \times 10^{14}\mathrm{cm}.
\end{equation}
Therefore, the inertial length is shorter than the wavelength of striped wind.
From the estimation described above, we can see that the small scale turbulence can survive in the wind region.

Next we consider the parameters for downstream.
We do not consider the possibility that the downstream plasma has a bulk relativistic speed.
The inertial length and Debye length are comparable at relativistic temperatures.
We adopt the value of typical Lorentz factor $ \gamma = 7 \times 10^3$ (Tanaka \& Takahara model),
and $\gamma = 10^6$ (Keneel \& Coroniti model).
Then we obtain the inertial length
\begin{equation}
\left(\frac{c}{\omega_{\mathrm{pe}}}\right)_\mathrm{d,TT} \sim 3 \times 10^{10}\mathrm{cm},\/
\end{equation}
\begin{equation}
\left(\frac{c}{\omega_{\mathrm{pe}}}\right)_\mathrm{d,KC} \sim 3 \times 10^{12}\mathrm{cm}.
\end{equation}
The wavelength of striped wind is compressed by a factor of a few $ \times \Gamma$ times 
compared to comoving wavelength in the upstream.
Therefore, the typical scale of magnetic field is 
\begin{equation}
\left(\lambda_{\mathrm{sw}}\right)_{\mathrm{d}} \sim 3 \times 10^8 \mathrm{cm}.
\end{equation}
From the estimation above, we see that the small scale turbulence decays far downstream.
We note that near the shock front or in the shock transition region, the plasma is not completely thermalized.
Therefore, the small scale turbulence can survive in some measure there.

Generally, when the Debye length is much larger than $\lambda_{\mathrm{sw}}$, the longitudinal mode would disappear rapidly.
However, when the dense blob enters the shock front, the inertial length becomes shorter and small scale turbulence tends to survive in longer time.
The density required for the survival far downstream is $10^5$ times larger than the mean density $n$, 
but even when the density contrast is less extreme, short wavelength turbulence required for the flares can exist in the shock transition region.

Summarizing this subsection, jitter radiation can produce the flare when the small scale turbulence survives in the shocked dense blob,
and the typical scale of turbulence is consistent with the typical frequency of the flares.
We propose the flare model that the high density blob plunge into the termination shock, an entropy mode is compressed or transformed to some other waveform 
in the shock transition region, the accelerated electrons move in this kind of turbulent field and radiate the highest energy photons by jitter mechanism.

\subsection{Energetics}
\begin{figure}
\includegraphics[width=12cm]{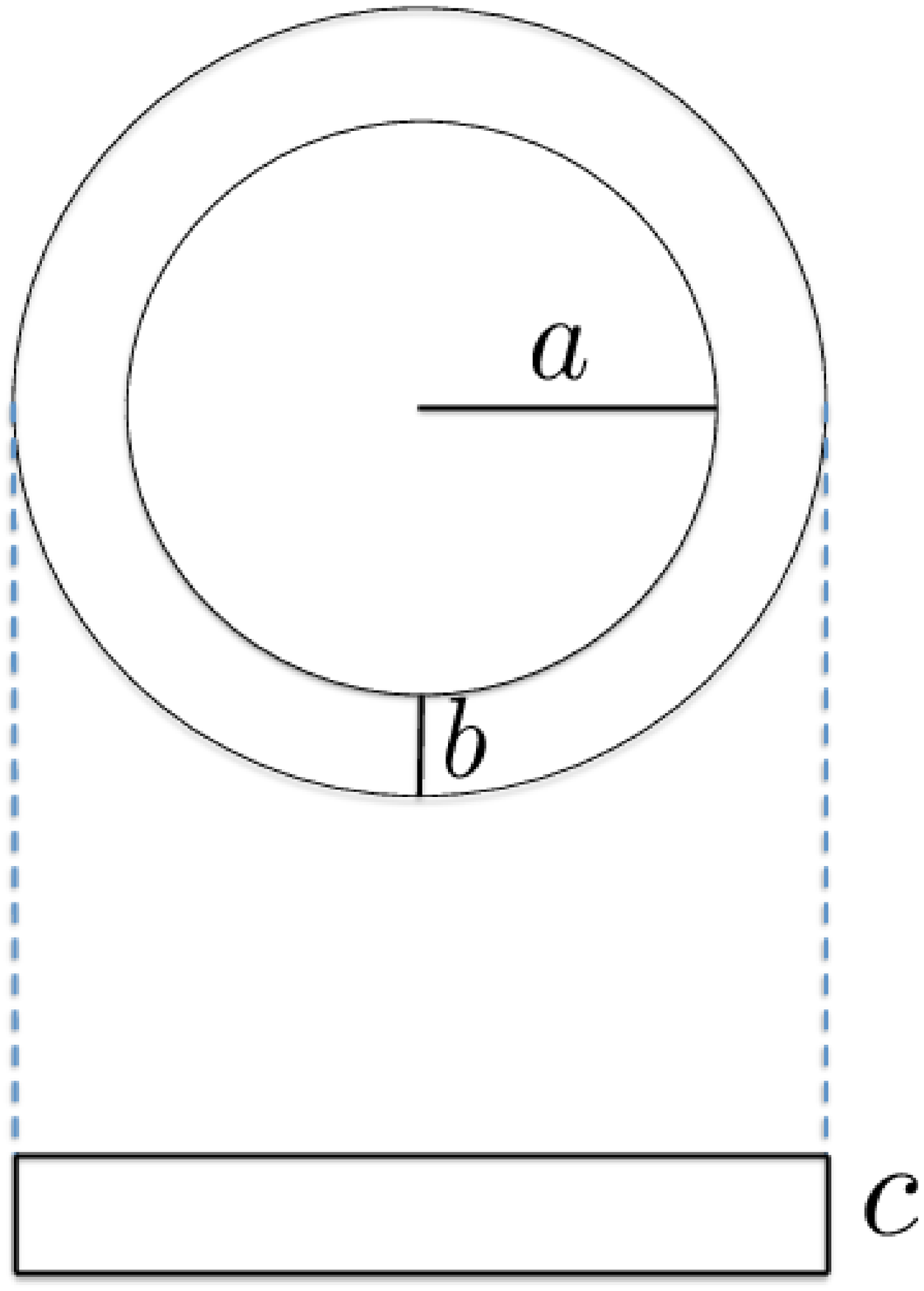}
\caption{Cartoon of the "100MeV ring".
The length $a$ is the radius of termination shock $r_{\mathrm{ts}} = 3\times 10^{17}\mathrm{cm}$,
$b$ is radial thickness and $c$ is the height of 100MeV ring. They are restricted by the Larmor radius of $3\times 10^{16}\mathrm{cm}$.
}
\label{er}
\end{figure}

Now that we have shown that the peak energy higher than $E_{\mathrm{c}}$ can be explained by jitter radiation, we next examine the energetics of flares.
Firstly we note that the energetics problem is very difficult to solve and has not been much addressed in previous models.
The scale of the emission region is constrained by the observed fluctuation time scale as
$c t_{\mathrm{fluc}} \sim 10^{15}\mathrm{cm}$ or by the duration timescale 
as $c t_{\mathrm{dur}} \sim 3 \times 10^{16}\mathrm{cm}$.
It is very difficult to concentrate $1\%$ of the spin down luminosity on this small region, compared to the circumference of the termination 
shock $\sim 2 \times 10^{18}\mathrm{cm}$, in either case.
We discuss the energetics by considering the size of the emission region and the density of radiating particles in it.
The Crab nebula is not spherically symmetric as is seen in the X-ray image by Chandra X-ray observatory (Gaensler \& Slane 2006).
It is possible that the emission regions of $100\mathrm{MeV}$ gamma-rays are patchy, but we do not resolve the Crab nebula at 100MeV gamma-rays, 
then we assume that the shape of the emission region is a ring as drawn in Fig. \ref{er}, for simplicity.
When the nebula is quiescent, the radial thickness is determined by synchrotron cooling.
To estimate the radial thickness, we suppose that the acceleration site is located only near the shock front, 
and the electrons return to the shock front on gyro time. 
If we assume the standard value of the strength of magnetic field $\mathrm{B} = 300 \mu \mathrm{G}$ (Kennel \& Coroniti 1984),
and considering the fact that cooling limits the attainable energy as $\gamma \sim 6 \times 10^{9}(\frac{B}{3 \times 10^{-4}\mathrm{G}})$,
we get the radial thickness of the ring as $r_\mathrm{L} \sim 3 \times 10^{16}\mathrm{cm}$.
When we assume $B = 85 \mu \mathrm{G}$ (Tanaka \& Takahara 2010), the thickness is three times larger.
We assume that the injection site of highest energy electrons is on the equatorial plane, so the ring height is also constrained by gyro radius of highest energy electrons.
The radius of the termination shock is $3\times 10^{17}\mathrm{cm}$, so the radial thickness and height of the 100MeV ring is a few $\times 10\%$ of the radius.

Next we estimate the parameters in the emission region in the flare state.
Firstly we examine the case when the scale of the blob is $c t_{\mathrm{fluc}} \sim 10^{15}\mathrm{cm}$, 
and the single blob becomes the emission region for the flare.
We assume that the blob moves on the equatorial plane, so a part of the ring becomes the emission region of flare. 
If we assume that the strength of magnetic field in the blob is the same as in the other region, 
the radial thickness of jitter emission region cannot be determined 
by synchrotron cooling, because the Larmor radius of the highest energy electrons 
$3\times 10^{16}\mathrm{cm} \left(\frac{B}{3\times 10^{-4}\mathrm{G}}\right)^{-3/2 }$ is larger than the blob size $ct_{\mathrm{fluc}}$.
The acceleration region is larger than the jitter emission region and the size of emission region is determined by blob size in this picture.
However, this picture does not work for flare models.
The reason is as follows.
The energy distribution of electrons at flare states is very hard and different from the one of the quiescent state.
Then the acceleration process in the acceleration region of the highest energy electrons which emit flare photons is different from other region.
We assume that a dense blob enters in the termination shock region, and implicitly assume that the other region is undisturbed.
Then the acceleration process outside the blob should be the same as in the quiescent state.
Therefore, it is more natural that the magnetic field in the blob is stronger than the mean magnetic field strength
and that the acceleration process is also different in the flare states to produce highest energy electrons with a very hard spectrum.
Thus, the cutoff energy of accelerated electrons should be smaller because of the strong magnetic field.
Since the size of acceleration region is limited by the blob size, 
the required strength of magnetic field is $3 \times 10^{-3}\mathrm{G}$ to make $r_{\mathrm{L}}=ct_{\mathrm{fluc}}$, 
and the maximum Lorentz factor is limited by radiation loss and becomes smaller to $\sim 2\times 10^9$.
Therefore, the required wavelength of turbulent field becomes $10^6\mathrm{cm}$ to emit $400 \mathrm{MeV}$ photons.
This constraint may seem to be very tight, but it is not improbable.
From this consideration, the volume of the blob is $10^{45}\mathrm{cm}^3$ and
the emission region of the flare is about $2 \times 10^{6}$ times smaller than in quiescent state,
because the volume of 100MeV ring is (circumference) $\times$ (radial thickness) $\times$ (height) $=2\times10^{51}\mathrm{cm}^3$.

\begin{figure}
\includegraphics[width=12cm]{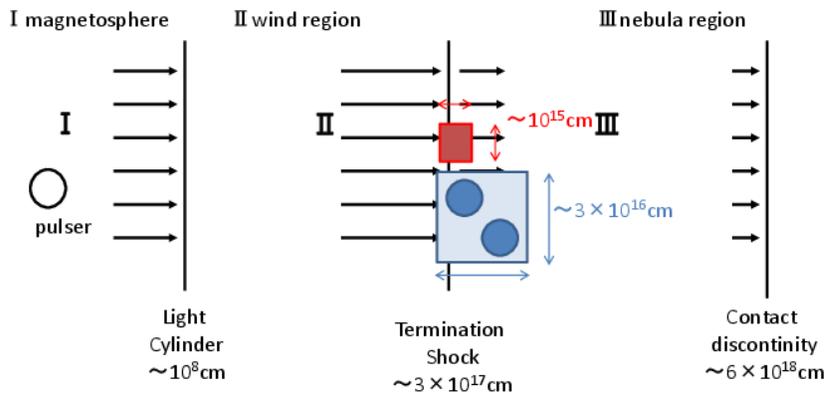}
\caption{A schematic picture for comparison of relevant scales.
The red box is the emission region of the homogeneous blob model, and the blue box is that of the inhomogeneous blob model.
}
\label{bl}
\end{figure}

The constraint for the volume of emission region can be alleviated when we assume the blob size is $ct_{\mathrm{dur}} = 3\times 10^{16}\mathrm{cm}$, 
and flux fluctuation comes from the internal structure of the blob of which scale is $ct_{\mathrm{fluc}} = 10^{15}\mathrm{cm}$. 
We assume that the acceleration scale is the same as the blob scale, and small denser regions
of which scale is $\sim ct_{\mathrm{fluc}}$ distribute in it as depicted in Fig. \ref{bl}.
The mean magnetic field strength is $3\times 10^{-4}\mathrm{G}$, by equating Larmor radius of highest energy electrons and $ct_{\mathrm{dur}}$.
The Lorentz factor of the highest energy electrons is determined by the magnetic field strength as $\gamma \sim 6 \times 10^{9}$,
and the required wavelength of turbulent field to emit $400$ MeV photon is estimated as $10^{7}\mathrm{cm}$.
The size of the blob is $3 \times 10^{16}\mathrm{cm}$,
which is the same as the thickness of the 100MeV ring in the quiescent state for the standard magnetic field strength.
Therefore the blob volume is only about $10^2$ times smaller than the 100MeV ring.

Next we consider the required number density of highest energy electrons in the blob to reproduce the flare luminosity.
We are considering the high density blob, so the number density of accelerated electrons can be much larger than the one of the quiescent state.
The luminosity is proportional to $\gamma^2 B^2 N$, where $N$ is the number of electrons at maximum energy in the blob.
First, for homogeneous blob of a size $ct_{\mathrm{fluc}}$, we assumed that the magnetic field strength is about $10$ times larger than the mean magnetic field strength.
Therefore the maximum Lorentz factor is limited as $2\times 10^9$, which is a few times smaller than the Lorentz factor of the highest energy electrons
in other regions in the 100MeV ring.
The volume of the emission region is $2 \times 10^{6}$ times smaller than that in a quiescent state.
Therefore, the required number density of the highest energy electrons in the blob is about $10^6$ times larger than in the quiescent state
to reproduce the flare luminosity.
In section \ref{sec:small}, we considered the required density for the survival of the small scale fluctuations in shock transition region.
It is about $10^5$ times the mean density.
If the acceleration is the same as in the quiescent state, the number density of the highest energy electrons may not be as large as $10^6$
times the number density of the highest energy electrons in the quiescent state.
However, the energy distribution of accelerated electrons is very hard, so the number of
the highest energy electrons can be $10^6$ times larger than in the quiescent state.
Therefore, the flare luminosity can be explained by this model if the mean density in blob fulfills the condition of the survival of the small scale turbulence.
Here, we have to note that the flare luminosity is $1\%$ of the spindown luminosity, so the asymmetry of the pulsar wind must be very high in this model.

Next, we examine the constraint on the scenario of inhomogeneous blob of a size $ct_{\mathrm{dur}}$.
The blob volume is only $10^2$ times smaller than the 100MeV ring, and we assumed that the magnetic field 
strength is the same order as the one of quiescent state ($3 \times 10^{-4}\mathrm{G}$), 
so the maximum Lorentz factor of the electrons is the same as in other region.
The required number density of highest energy electrons in the blob is about $10^2 - 10^3$ times larger than the mean density of highest energy electrons.
The flare luminosity can be obtained by considering the hardness of electron energy distribution which is calculated from the observed flux alone, 
and the high number density of electrons would help to accomplish the large luminosity of flare.
In short, while the small homogeneous blob scenario is not impossible, large inhomogeneous blob scenario is more plausible.

\subsection{Spectrum}
The observed spectra of flares indicate that the energy distribution of electrons is very hard.
As is discussed in the previous subsection, the hard energy distribution of electrons is also required to solve the energetics.
If the electrons take a power law energy distribution, the power law index $p$ of electrons ($\frac{dN}{dE} \propto E^{-p}$) can be estimated from the photon index.
However, when the strength parameter $a<1$ and when either $p<1$ or $p<2\mu+1$, the photon index around 100MeV can be determined by jitter mechanism,
where $\mu$ is the power law index of isotropic turbulent magnetic field ($B^2(k)\propto k^{-\mu}$).
The spectrum of flare component is fitted by a power law plus cutoff,
and the time integrated power law index is $\gamma_{\mathrm{F}} = 1.27 \pm 0.12$ (Buehler et al. 2011).
Clausen-Brown \& Lyutikov reexamined the time resolved spectrum in Buheler et al. 2012, 
and obtained the photon index in the most luminous period as
\begin{equation}
\gamma_{\mathrm{F}} = 1.08 \pm 0.16.
\end{equation}
If the index is supposed to reflect the energy distribution of electrons, the time integrated
power law index is $p = 1.54 \pm 0.24$ and time resolved one (in the most luminous state) is $p = 1.16 \pm 0.32$, because 
$\gamma_{\mathrm{F}} = (p+1)/2$.
It is very hard and inconsistent with the power law index $p = 2.5$ at injection in the quiescent state (Tanaka \& Takahara 2010).
Additionally, the hard energy distribution is consistent with the observation that no counterpart of the flares has been detected in other wavelengths.
From these facts, the particle acceleration in the blob is expected to be different from the other region.
For example, a stochastic acceleration process may play a crucial role to make the hard electron energy distribution in a short time (see e.g. Hoshino 2012).

\begin{figure}
\includegraphics[width=12cm]{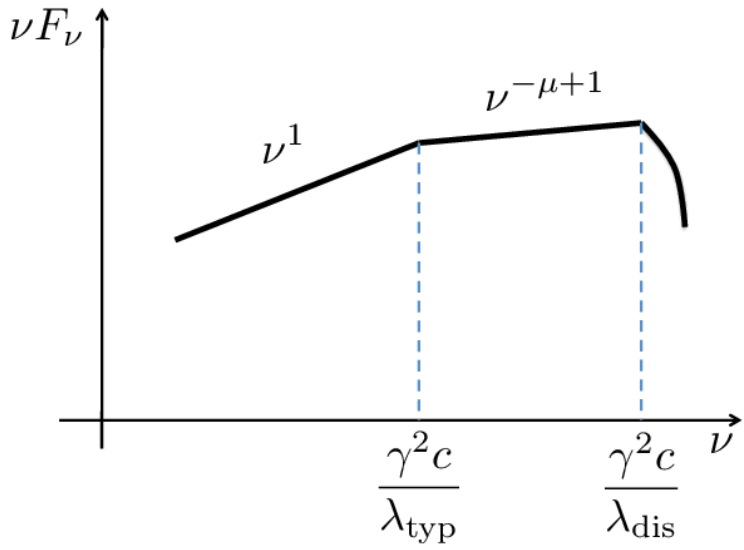}
\caption{Radiation spectrum of jitter radiation by monoenergetic electrons for $a<1$.
}
\label{sp}
\end{figure}

The hard photon index can be interpreted as the reflection of hard power law index of electron energy distribution, 
but getting the value $p\sim 1$ is somewhat difficult (Clausen-Brown \& Lyutikov 2012).
We show another interpretation of these spectral indices by using the theory of jitter radiation on the assumption that 
the accelerated electrons follow a very hard, almost monoenergetic distribution.
For $a<1$, the theoretical spectrum of jitter radiation from monoenergetic particles moving in an isotropic turbulent magnetic field
is expressed as a broken power law and cutoff as is seen in Fig.\ref{sp} (e.g. Fleishman 2006).
The photon index of the low energy side is 
\begin{equation}
\gamma_{\mathrm{F}} = 1,
\end{equation}
and that of the high energy side is $\gamma_{\mathrm{F}} = \mu +1$.
The cutoff energy is determined by the smallest scale (in other words, dissipation scale $\lambda_{\mathrm{dis}}$) of the turbulent field.
The inertial length, which corresponds to the typical scale of magnetic field fluctuations, is proportional to $n^{1/2}$, and the luminosity is proportional to $n$ 
when the emission region volume is fixed.
Therefore, the typical photon energy of flares is the highest in the most luminous state.
Additionally, the typical energy and flux should have the positive correlation in this model, and it is consistent with the observation (Buehler et al. 2011). 
We regard the photon index of $\gamma_{\mathrm{F}} = 1.08 \pm 0.16$ as the intrinsic photon index of jitter radiation.
At this time, the dissipation scale $\lambda_{\mathrm{dis}}$ and injection scale $\lambda_{\mathrm{typ}}$ of turbulent field are very close, 
so it is difficult to resolve cutoff frequency  $\gamma^2 c/\lambda_{\mathrm{dis}}$ and break frequency $\gamma^2 c/ \lambda_{\mathrm{typ}}$. 
When the flux is smaller, the injection scale would be larger because the number density would be smaller. 
Therefore, $\gamma^2 c/\lambda_{\mathrm{typ}}$ becomes smaller, 
so the photon index around 100MeV can be interpreted as a reflection of the power law index of magnetic field fluctuations.
It should be noted that it is usual $\mu>1$ in the ordinary turbulent field, which causes some problem that $\mu<1$ is required to  
explain the observed spectral index.

\section{Discussion }
\subsection{The difference from other models and predictions}
We have considered inhomogeneities of the emission region.
Bykov et al. also considered inhomogeneous emission regions.
At first, we discuss the difference from their model.
They assumed that the size of the acceleration site is much larger than the emission region, and 
the acceleration mechanism in the quiescent state and flare state is identical.
If the energy distribution of electrons stays unchanged in the flare state, the spectra in 100MeV range cannot become harder 
than the spectrum in MeV range in the quiescent state.
This does not seem to match observations.
In contrast, we consider that the acceleration site should have a similar size to the blob size, and the acceleration mechanism is different in the flares,
because the observed spectrum is very hard.
When the electron energy distribution is very hard ($p\leq1$), 
the photon index $\gamma_{\mathrm{F}} = 1$ can be naturally explained by jitter mechanism.
They deal with the problem assuming that the emission region is 1D for radial direction, and they do not consider the energetics explicitly in their paper.
They consider the radial length of the emission region is the same as the quiescent state ($ 2 \times 10^{16} \mathrm{cm}$),
and there are the blobs randomly distributed with the $1\%$ scale ($\sim 10^{14}\mathrm{cm}$) having stronger magnetic field.
The length is consistent with the observed timescale of the flares.
The scale corresponding to the single pulse of flare has to be smaller than $10^{15}\mathrm{cm}$.
The solid angle of the emission region can be constrained by duration time of flare.
Therefore, the predicted luminosity is a few dozen times smaller than the observed one.
While they predicted that the polarization degree would enhance during the flare,
our model predicts the converse prediction. The polarization degree would be very low during the flare, 
because the gamma-rays are emitted in the turbulent field by jitter mechanism.

The most popular interpretations of the Crab flares are Doppler boost models.
While Doppler boost model predicts that the TeV-PeV flare would accompany the 100MeV flare,
 our model does not predict such a correspondence between GeV and TeV-PeV.
In our model, the increase of the highest energy electrons and frequency shift collaborate to create the flare.
Therefore, the required total number of the highest energy electrons is only a few times larger than the quiescent state.
In TeV-PeV range, since there are no frequency shift, and inverse Compton scattering by the highest energy electrons are in Klein-Nishina regime
so that only a very weak bump will appear in PeV range.

The hard spectrum of flares is one of the difficult features to interpret.
Clausen-Brown \& Lyutikov explained this hard spectrum by very hard electron energy distribution near the radiation reaction limit.
They assumed acceleration time much shorter than escaping time, and considered radiation loss.
The electrons pile up near the maximum energy.
They commented that the pile-up scenario could explain the observed SED by tuning the acceleration timescale.
If acceleration time is much shorter than the fluctuation time of flare, the distribution becomes monoenergetic,
and spectrum becomes intrinsic one $\gamma_{\mathrm{F}} = 2/3$ for synchrotron radiation or $\gamma_{\mathrm{F}}= 1$ for jitter radiation.
Our model does not require the tuning of acceleration time, and predict that the flare spectrum will not be harder than $\nu F_{\nu}\propto \nu^1$.

\subsection{The acceleration and scatterers}
Kirk and Reville argued that jitter radiation cannot emit photons with energy higher than the critical synchrotron energy 
in the DSA scenario in their paper (Kirk \& Reville 2010).
In their analysis, they assumed that the scatterer (magnetic field fluctuation) is a single population.
For $a<1$, particles experience ballistic transport and take a longer time to come back to the shock than the gyrotime.
Therefore, the acceleration time becomes longer, so the maximum energy of electrons becomes smaller,
and radiation frequency is smaller than the one for $a>1$ in spite of taking into account jitter mechanism.
Conversely we argue that the jitter mechanism can emit higher energy radiation than synchrotron one.
The reason for apparently inconsistent conclusions lies in the difference of situations.
We assumed implicitly the existence of multi populations of scatterers.
Although we do not specify the acceleration mechanism, we suppose that the large scale scatterers exist, too.
The acceleration time depends on the large scale (as large as Larmor radius) scatterers, so the acceleration time is not so long.
Therefore, our model does not contradict their conclusion.
In fact, the situation with two populations of scatterers are considered by Reville \& Kirk 2010, and jitter component emerges over the synchrotron cut off.
Stated another way, the photon energy of jitter component can be higher than that of synchrotron component when there are multi population of scatterers.

\section{Summary \& Conclusion}
We propose a model which explains the flares of the Crab nebula over the 100MeV by jitter radiation.
The wavelength of striped wind of the Crab pulsar is about two order of magnitude longer than the required scale of turbulent field 
to emit photons with energy $E>E_{\mathrm{c}}$ by jitter mechanism.
A high density region is required for existence of the small scale turbulence.
It is suggested that there are large density fluctuations in the Crab pulsar magnetosphere and nebula.
Therefore, we consider that there are high density blobs in the pulsar wind region.
The blobs plunge into the termination shock, generate the short wavelength turbulence of electromagnetic field,
and accelerated electrons radiate gamma-ray emission by jitter mechanism in the blob.
The required strength of mean magnetic field in blob is $10$ times larger, and the number density of highest energy electrons in blob is
$10^6$ times larger than in quiescent state to reproduce the April 2011 flare by homogeneous blob model
for which the size of the blob is $ct_{\mathrm{fluc}}\sim 10^{15}\mathrm{cm}$.
When we adopt the inhomogeneous blob model, for which the size of the blob is $ct_{\mathrm{dur}}\sim 3\times 10^{16}\mathrm{cm}$, 
the required magnetic field strength is as large as that of the quiescent one, and
number density of highest energy electrons is about $10^2-10^3$ times larger than in the quiescent state.
The required high density of highest energy electrons in the blob is consistent with our assumption that high density blobs trigger flares and
hard energy distribution of electrons which is implied by observed spectra.
The very hard photon index $\gamma_{\mathrm{F}} =1.08 \pm 0.16$ of April $2011$ flare in the brightest state is
consistent with the intrinsic photon index of jitter radiation for $a<1$.
We make following three predictions for the future Crab flares:
firstly, the polarization degree will become lower in flare state, secondly, no counterpart will be seen in TeV-PeV range, and 
thirdly, the flare spectrum will not be harder than $\nu F_{\nu}\propto \nu^1$.

We thank the referee for helpful comments.
We are grateful to Y. Ohira for useful discussions.
This work is supported by JSPS Reserch Fellowships for Young Scientists (Y. T.,24593).

\clearpage

\end{document}